# VARIATIONAL STRUCTURE OF THE OPTIMAL ARTIFICIAL DIFFUSION METHOD FOR THE ADVECTION-DIFFUSION EQUATION

K. B. NAKSHATRALA AND A. J. VALOCCHI

Abstract. In this research note, we provide a variational basis for the optimal artificial diffusion method, which has been a cornerstone in developing many stabilized methods. The optimal artificial diffusion method produces exact nodal solutions when applied to one-dimensional problems with constant coefficients and forcing function. We first present a variational principle for a multi-dimensional advective-diffusive system, and then derive a new stable weak formulation. When applied to one-dimensional problems with constant coefficients and forcing function, this resulting weak formulation will be equivalent to the optimal artificial diffusion method. We present representative numerical results to corroborate our theoretical findings.

## 1. INTRODUCTION

Many transport-related processes are modeled as advective-diffusive system. For example, transport of contaminants in subsurface flows is modeled as an advection-diffusion equation coupled with Darcy flow. Except for very simple and limited problems, one cannot find analytical solutions and hence must resort to numerical solutions. However, it is well-known that great care should be taken in developing numerical formulations in order to avoid spurious oscillations due to the advective term.

Many numerical formulations have been proposed, which fall under the realm of stabilized methods. One such method is the optimal artificial diffusion method, which has been the basis for developing many stabilized methods and has also served as a benchmark for comparison [6, 2]. The optimal artificial diffusion method is derived by imposing the condition that it should produce exact nodal solutions for one-dimensional problems with constant coefficients and forcing function. This derivation, as one can see, does not have a variational basis.

Herein we outline a variational structure behind the optimal artificial diffusion method. We start with a variational statement for the advection-diffusion equation (which is not a self-adjoint operator), and derive a stable weak formulation. The resulting weak formulation when applied to one-dimensional problems with constant coefficients and forcing function produces the same difference equation as the optimal artificial diffusion method, which produces exact nodal solutions when applied to such problems. This shows that the optimal artificial diffusion method has a firm variational basis. This paper also highlights other possible routes in developing stable formulations for non-self-adjoint operators.

An outline of this short paper is as follows. In Section 2, we present governing equations for an advective-diffusive system, and also outline the standard weighted residual method. In Section 3, we describe the optimal artificial diffusion method. In Section 4, we briefly discuss Vainberg's theorem, which provides a connection between the weighted residual statement and its corresponding scalar functional (if it exists). In Section 5, we present a variational principle for the advection-diffusion equation, and then derive a stable weak formulation. We then show that the resulting weak formulation produces the same difference equation as the optimal artificial diffusion method for one-dimensional problems with constant coefficients and forcing function. Finally, we draw conclusions in Section 6.

**Remark 1.1.** *It should be noted that there is a huge literature on developing stabilized (finite element) formulations for an advective-diffusive system. For example, see [1, 3, 12, 4] and references therein. A thorough discussion of these works is beyond the scope of this paper. In addition, these discussions are not relevant to the subject matter of this paper as none of them discuss a variational principle (that is, constructing a scalar functional) for an advection-diffusion system.*

## 2. GOVERNING EQUATIONS: ADVECTIVE-DIFFUSIVE SYSTEM

Let $\Omega \subset \mathbb{R}^d$ be a smooth and bounded domain, where "$d$" denotes the number of spatial dimensions, and $\partial \Omega$ denotes its smooth boundary. As usual, the boundary is divided into $\Gamma^\mathrm{D}$ (the part of the boundary on which Dirichlet boundary condition is prescribed) and $\Gamma^\mathrm{N}$ (the part of the boundary on which Neumann boundary condition is prescribed) such that $\Gamma^\mathrm{D} \cup \Gamma^\mathrm{N} = \partial \Omega$ and $\Gamma^\mathrm{D} \cap \Gamma^\mathrm{N} = \emptyset$. Let $\boldsymbol{x} \in \Omega$ denote the position vector, and the gradient and divergence operators are denoted as "grad" and "div", respectively. Let $u : \Omega \rightarrow \mathbb{R}$ denote the concentration, $\boldsymbol{v}(\boldsymbol{x})$ the velocity vector field, and $\boldsymbol{k}(\boldsymbol{x})$ the symmetric and positive-definite diffusivity tensor. For further discussion consider the following steady advective-diffusive system

$$\boldsymbol{v}(\boldsymbol{x}) \cdot \mathrm{grad}[u(\boldsymbol{x})] - \mathrm{div}\,[\boldsymbol{k}(\boldsymbol{x})\,\mathrm{grad}[u(\boldsymbol{x})]] = f(\boldsymbol{x}) \quad \text{in } \Omega \tag{2.1}$$

$$u(\boldsymbol{x}) = u^\mathrm{p}(\boldsymbol{x}) \quad \text{on } \Gamma^\mathrm{D} \tag{2.2}$$

$$\boldsymbol{n}(\boldsymbol{x}) \cdot \boldsymbol{k}(\boldsymbol{x}) \mathrm{grad}[u(\boldsymbol{x})] = t^\mathrm{p}(\boldsymbol{x}) \quad \text{on } \Gamma^\mathrm{N} \tag{2.3}$$

where $u^\mathrm{p}(\boldsymbol{x})$ is the prescribed Dirichlet boundary condition, $t^\mathrm{p}(\boldsymbol{x})$ is the prescribed Neumann boundary condition, $f(\boldsymbol{x})$ is the prescribed volumetric source, and $\boldsymbol{n}(\boldsymbol{x})$ denotes the unit outward normal to the boundary. Note that the advective-diffusive operator is *not* self-adjoint. It is well-known that, even under smooth functions for $\boldsymbol{k}(\boldsymbol{x}), \boldsymbol{v}(\boldsymbol{x})$ and $f(\boldsymbol{x})$, the solution $u(\boldsymbol{x})$ to equations (2.1)–(2.3) may exhibit steep gradients close to the boundary especially for advection-dominated problems [6] (and also see Figure 1, which will be described later in Section 3).

**2.1. Notation and preliminaries.** In the next section, we present a weak formulation for an advective-diffusive system. To this end, let us define the following function spaces:

$$\mathcal{U} := \left\{ u(\boldsymbol{x}) \in H^1(\Omega) \mid u(\boldsymbol{x}) = u^{\mathrm{p}}(\boldsymbol{x}) \text{ on } \Gamma^{\mathrm{D}} \right\} \tag{2.4a}$$

$$\mathcal{W} := \left\{ w(\boldsymbol{x}) \in H^1(\Omega) \mid w(\boldsymbol{x}) = 0 \text{ on } \Gamma^{\mathrm{D}} \right\} \tag{2.4b}$$

where $H^1(\Omega)$ is a standard Sobolev space defined on $\Omega$. Note that the inner product for the above vector spaces is the standard $L^2$ inner product. That is,

$$(a;b) := \int_\Omega a(\boldsymbol{x}) \cdot b(\boldsymbol{x}) \, \mathrm{d}\Omega \tag{2.5}$$

Similarly, one can define the weighted $L^2$ inner product

$$(a;b)_\mu := \int_\Omega \mu(\boldsymbol{x}) a(\boldsymbol{x}) \cdot b(\boldsymbol{x}) \, \mathrm{d}\Omega \tag{2.6}$$

where $\mu : \Omega \to \mathbb{R}^+$ (where $\mathbb{R}^+$ denotes the set of positive real numbers) is the (scalar) weight function or measure density. Note that a weight function (which is used to define the weighted inner product) should not be confused with weighting functions (which are sometimes referred to as test functions). Using this weighted inner product one can define weighted Sobolev spaces as

$$L^2_\mu(\Omega) := \left\{ u(\boldsymbol{x}) \mid (u;u)_\mu < +\infty \right\} \tag{2.7a}$$

$$H^1_\mu(\Omega) := \left\{ u(\boldsymbol{x}) \in L^2_\mu(\Omega) \mid (\mathrm{grad}[u]; \mathrm{grad}[u])_\mu < +\infty \right\} \tag{2.7b}$$

Corresponding to the function spaces given in (2.4), we can define the following weighted function spaces

$$\mathcal{U}_\mu := \left\{ u(\boldsymbol{x}) \in H^1_\mu(\Omega) \mid u(\boldsymbol{x}) = u^{\mathrm{p}}(\boldsymbol{x}) \text{ on } \Gamma^{\mathrm{D}} \right\} \tag{2.8a}$$

$$\mathcal{W}_\mu := \left\{ w(\boldsymbol{x}) \in H^1_\mu(\Omega) \mid w(\boldsymbol{x}) = 0 \text{ on } \Gamma^{\mathrm{D}} \right\} \tag{2.8b}$$

**2.2. Standard weighted residual method and the Galerkin formulation.** Let $w(\boldsymbol{x})$ denote the weighting function corresponding to $u(\boldsymbol{x})$. A weak formulation based on the *standard* weighted residual method for the advective-diffusive system given by equations (2.1)-(2.3) can be written as

$$\text{Find } u(\boldsymbol{x}) \in \mathcal{U} \text{ such that } \mathcal{F}(w;u) = 0 \ \forall w(\boldsymbol{x}) \in \mathcal{W} \tag{2.9}$$

where the bilinear functional $\mathcal{F}$ is defined as

$$\mathcal{F}(w;u) := \int_\Omega w(\boldsymbol{x}) \boldsymbol{v}(\boldsymbol{x}) \cdot \mathrm{grad}\,[u(\boldsymbol{x})] \, \mathrm{d}\Omega + \int_\Omega \mathrm{grad}\,[w(\boldsymbol{x})] \cdot \boldsymbol{k}(\boldsymbol{x}) \mathrm{grad}\,[u(\boldsymbol{x})] \, \mathrm{d}\Omega$$
$$- \int_\Omega w(\boldsymbol{x}) f(\boldsymbol{x}) \, \mathrm{d}\Omega - \int_{\Gamma^{\mathrm{N}}} w(\boldsymbol{x}) t^{\mathrm{p}}(\boldsymbol{x}) \, \mathrm{d}\Gamma \tag{2.10}$$

A finite element formulation corresponding to the standard weighted residual formulation can be written as

(2.11) $$\text{Find } u^h(\boldsymbol{x}) \in \mathcal{U}^h \text{ such that } \mathcal{F}(w^h; u^h) = 0 \; \forall w^h(\boldsymbol{x}) \in \mathcal{W}^h$$

where $\mathcal{U}^h$ and $\mathcal{W}^h$ are finite dimensional subspaces of $\mathcal{U}$ and $\mathcal{W}$, respectively. One obtains the Galerkin formulation by using the same function space for both $\mathcal{U}^h$ and $\mathcal{W}^h$ (except on $\Gamma^{\mathrm{D}}$).

The Galerkin formulation produces node-to-node spurious oscillations for advection-dominated problems [6]. The Galerkin method loses its best approximation property when the non-symmetric convective operator dominates in the transport equation (for example, see the performance of the Galerkin formulation in Figure 1). In other words, the Galerkin method is *not* optimal for solving advection-dominated problems. In principle, it is possible to choose a small enough grid such that the element Péclet number is less than one, and avoid spurious oscillations under the Galerkin method. However, it may not always be practical to choose such a fine grid, and therefore, one needs to employ a stabilized formulation to avoid unphysical oscillations and get meaningful results on coarse grids. To understand this anomalous behavior several theoretical and numerical studies have been performed, see [4, 6, 2] and references therein. One of those studies is a simple method that gives nodally exact solutions for one-dimensional problems with constant coefficients, which is commonly referred to as the optimal diffusion method.

## 3. A NODALLY EXACT FORMULATION AND OPTIMAL ARTIFICIAL DIFFUSION

Consider the following one-dimensional advection-diffusion equation with homogeneous Dirichlet boundary conditions, and constant coefficients (that is, velocity, diffusivity and forcing function are constants):

(3.1) $$v\frac{du}{dx} - k\frac{d^2u}{dx^2} = f \quad \forall x \in (0,1)$$

(3.2) $$u(x=0) = 0 \text{ and } u(x=1) = 0$$

The analytical solution for the above problem is given by

(3.3) $$u(x) = \frac{f}{v}\left(x - \frac{1-\exp(vx/k)}{1-\exp(v/k)}\right)$$

which is plotted in Figure 1 for various value of $v/k$. As one can see, for large $v/k$ we have steep gradients near the outflow boundary. The above problem has been used as a benchmark for developing many stabilized formulations [6].

For a numerical solution, let us divide the unit interval into $N$ equal-sized elements (and hence $N+1$ nodes), and define $h := 1/N$. Let the nodes be numbered as $j = 0, \cdots, N$. Then, the position vector of node $j$ is $x_j = jh$. The difference equation at an intermediate node ($j = 1, \cdots, N-1$) arising from the Galerkin formulation for the above problem can be written as

$$\text{(3.4)} \qquad \frac{v}{2h}\left(-\frac{P_e^h+1}{P_e^h}u_{j-1} + \frac{2}{P_e^h}u_j + \frac{P_e^h-1}{P_e^h}u_{j+1}\right) = f$$

where $P_e^h = vh/(2k)$ is the element Péclet number, and $u_j$ is an approximate numerical solution at node $j$. That is,

$$\text{(3.5)} \qquad u_j \approx u(x_j) \quad \forall j = 0, \cdots, N$$

The difference equation (3.4) can be rearranged as

$$\text{(3.6)} \qquad v\frac{u_{j+1} - u_{j-1}}{2h} - k\frac{u_{j+1} - 2u_j + u_{j-1}}{h^2} = f$$

The above equation basically reveals that the Galerkin formulation approximates the first- and second-derivatives using a central difference approximation at intermediate nodes.

It is well-known that the Galerkin formulation is under-diffusive when applied to the advection-diffusion equation, which is considered to be the reason why the formulation gives node-to-node spurious oscillations. In Figure 1, we compare the numerical solution from the Galerkin formulation with the analytical solution for the aforementioned one-dimensional problem. As one can see from the figure, the Galerkin formulation produces spurious node-to-node oscillations for high Péclet numbers.

To also understand this anomalous behavior of the Galerkin formulation, we now outline a nodally exact formulation for the above problem, which is commonly referred to as the optimal artificial diffusion formulation in the literature. To this end, we start with a difference equation at node $j$ of the form

$$\text{(3.7)} \qquad \beta_{-1}u_{j-1} + \beta_0 u_j + \beta_1 u_{j+1} = f$$

and the unknown coefficients ($\beta_{-1}$, $\beta_0$, and $\beta_1$) are determined so that the above difference equation gives nodally exact solution for the model problem given by equations (3.1)-(3.2) on the uniform computational mesh described above. After simplification, the difference equation for the nodally exact formulation can be written as: (see References [2, 6, 16]),

$$\text{(3.8)} \qquad \frac{v}{2h}\left[-(1 + \coth(P_e^h))u_{j-1} + (2\coth(P_e^h))u_j + (1 - \coth(P_e^h))u_{j+1}\right] = f$$

where "coth" denotes the hyperbolic cotangent function. By rearranging the terms one can write the above equation as follows

$$\text{(3.9)} \qquad v\frac{u_{j+1} - u_{j-1}}{2h} - (k + \bar{k})\frac{u_{j+1} - 2u_j + u_{j-1}}{h^2} = f$$

where the artificial diffusion coefficient $\bar{k}$ (which depends on the mesh size, and medium and flow properties) is given by

$$(3.10) \qquad \bar{k} := \frac{vh}{2}\left(\coth(P_e^h) - \frac{1}{P_e^h}\right)$$

The difference equation (3.9) reveals that the nodally exact formulation (or the optimal artificial diffusion method) also employs a central difference approximation for the first- and second-derivatives but solves a modified advective-diffusive system with an additional artificial diffusion given by the coefficient $\bar{k}$.

It should be noted that the above derivation for the optimal artificial diffusion method is not based on a variational principle. In this paper we present a variational structure behind the optimal artificial diffusion method, which, to the best of our knowledge, has not been reported in the literature. We start by presenting a variational principle for an advective-diffusive system and then derive a weak formulation. A finite element approximation of this weak form gives the optimal artificial diffusion method.

We now present a variational principle for an advective-diffusive system. We then outline the underlying variational structure behind the optimal artificial diffusion method.

## 4. VAINBERG'S THEOREM AND EXISTENCE OF A SCALAR FUNCTIONAL

Vainberg's theorem [15] provides a connection between scalar functionals (also referred to as "energy" functionals) and weighted residual statements. The theorem provides a criterion to establish when a scalar functional exists. The theorem also provides a formula to compute the scalar functional (if it exists) from the weighted residual statement, which we will not invoke in this paper.

Vainberg's theorem can be stated as follows. Let $\mathcal{G}(w; u)$ be a weighted residual functional, which is linear with respect to $w$ (but need not be linear with respect to $u$). There exists a scalar functional $\mathcal{E}(u)$ such that

$$\mathcal{G}(w; u) = \left[\frac{d}{d\epsilon}\mathcal{E}(u + \epsilon w)\right]_{\epsilon=0}$$

if and only if

$$(4.1) \qquad \left[\frac{d}{d\epsilon}\mathcal{G}(w, u + \epsilon\bar{w})\right]_{\epsilon=0} = \left[\frac{d}{d\epsilon}\mathcal{G}(\bar{w}, u + \epsilon w)\right]_{\epsilon=0}$$

For a simple proof of Vainberg's theorem see Reference [9, Chapter 9].

Returning back to our discussion on advective-diffusive system, using Vainberg's theorem one can conclude that there exists *no* scalar functional $\mathcal{E}(u)$ whose directional derivative along $w(\boldsymbol{x})$ gives the (bilinear) functional in the standard weighted residual method (which is given by equation

(2.10)). That is, there is *no* scalar functional $\mathcal{E}(u)$ such that

$$\mathcal{F}(w; u) = \left[\frac{d}{d\epsilon} \mathcal{E}(u + \epsilon w)\right]_{\epsilon=0}$$

However, we should note conclude that the advective-diffusive system does not have a variational principle. All we concluded above (invoking Vainberg's theorem) is that the standard weighted residual statement (2.10) cannot be obtained as a directional derivative of a scalar functional. It may be possible to construct a different weighted residual statement with different weight function (or measure) that can be obtained as a directional derivative of a scalar functional. Stated differently, just because the operator is non-self-adjoint does not mean that there exists no variational statement.

In a seminal paper, Tonti [14] has correctly highlighted the point that the symmetry of a bilinear form in the sense of equation (4.1) (which according to Vainberg's theorem is a necessary and sufficient condition for the existence of a variational formulation) depends on the choice of the inner product. (Note that Tonti has used the acronym "bilinear functional" to mean inner product.) One of the ways to meet the symmetry requirement is by changing the underlying inner product. Tonti [14] has also shown that changing the inner product is equivalent to transforming the given problem into a different problem that has the same solution(s). Another related work is by Thangaraj and Venkatarangan [13] who have presented dual variational principles with applications to magneto-hydrodynamics. A related and important work is by Magri [11] who has developed a procedure to select an appropriate inner product for any given linear operator to meet the symmetry requirement. Following the discussion in References [14, 13], in the next section we present a variational principle for an advection-diffusion system.

**Remark 4.1.** *For self-adjoint operators (e.g., the Laplacian operator) one can show that the standard weighted residual method can be obtained as a directional derivative of a scalar functional.*

## 5. A VARIATIONAL PRINCIPLE FOR THE ADVECTION-DIFFUSION EQUATION

Consider the following constrained extremization problem

$$(5.1) \qquad \underset{u(\boldsymbol{x})}{\text{extremize}} \int_\Omega F\left(\boldsymbol{x}, u(\boldsymbol{x}), \text{grad}[u(\boldsymbol{x})]\right) \mathrm{d}\Omega$$

where $F : \Omega \times \mathbb{R} \times \mathbb{R}^d \to \mathbb{R}$ is a known scalar functional. It is well-known (for example, see [5, 10]) that the Euler-Lagrange equation for the above extremization problem is given by

$$(5.2) \qquad F_u\left(\boldsymbol{x}, u(\boldsymbol{x}), \text{grad}[u(\boldsymbol{x})]\right) - \text{div}\left[F_{\boldsymbol{p}}(\boldsymbol{x}, u(\boldsymbol{x}), \text{grad}[u(\boldsymbol{x})])\right] = 0 \quad \forall \boldsymbol{x} \in \Omega$$

where $F_u$ denotes the partial derivative with respect to $u(\boldsymbol{x})$, and $F_{\boldsymbol{p}}$ denotes the vector of partial derivatives with respect to the components of $\text{grad}[u(\boldsymbol{x})]$.

We now construct a scalar functional for the advection-diffusion equation. To this end, let us define the scalar function $\alpha(\boldsymbol{x})$ as

$$\alpha(\boldsymbol{x}) := \exp\left[-\int \left(\boldsymbol{k}^{-1}(\boldsymbol{x})\boldsymbol{v}(\boldsymbol{x})\right) \cdot \mathrm{d}\boldsymbol{x}\right] \tag{5.3}$$

Note that the integral inside the exponential operator in equation (5.3) is an *indefinite line integral*. Since the tensor $\boldsymbol{k}(\boldsymbol{x})$ is positive-definite, its inverse exists, and hence the scalar function $\alpha(\boldsymbol{x})$ is well-defined. It is important to note that $\alpha(\boldsymbol{x}) > 0 \;\forall \boldsymbol{x} \in \Omega$ (as we have assumed the domain to be bounded and smooth). Now define the functional for the advection-diffusion equation as

$$F(\boldsymbol{x}, u(\boldsymbol{x}), \mathrm{grad}[u(\boldsymbol{x})]) := \alpha(\boldsymbol{x}) \left(\frac{1}{2} \|\sqrt{\boldsymbol{k}(\boldsymbol{x})}\mathrm{grad}\,[u(\boldsymbol{x})]\|^2 - u(\boldsymbol{x})f(\boldsymbol{x})\right) \quad \forall \boldsymbol{x} \in \Omega \tag{5.4}$$

where $\|\cdot\|$ denotes the standard Euclidean norm, and $\sqrt{\boldsymbol{k}(\boldsymbol{x})}$ denotes the square root of the symmetric and positive-definite tensor $\boldsymbol{k}(\boldsymbol{x})$. (Note that the square root of a symmetric positive-definite tensor is well-defined, and is also symmetric and positive-definite. See Halmos [8] and also Gurtin [7, page 13].) A straightforward calculation shows that the Euler-Lagrange equation of the functional (5.4) is

$$\boldsymbol{v}(\boldsymbol{x}) \cdot \mathrm{grad}[u(\boldsymbol{x})] - \mathrm{div}\,[\boldsymbol{k}(\boldsymbol{x})\,\mathrm{grad}[u(\boldsymbol{x})]] = f(\boldsymbol{x}) \tag{5.5}$$

which is the (multi-dimensional) advection-diffusion equation. For the advective-diffusive system given by equations (2.1)-(2.3) (that is, including the boundary conditions), the variational statement can be written as

$$\underset{u(\boldsymbol{x})}{\mathrm{extremize}} \quad \mathcal{I}(u) \tag{5.6}$$

$$\mathrm{subject\ to} \quad u(\boldsymbol{x}) = u^{\mathrm{p}}(\boldsymbol{x}) \text{ on } \Gamma^{\mathrm{D}} \tag{5.7}$$

where the functional $\mathcal{I}(u)$ is defined as

$$\mathcal{I}(u) := \int_{\Omega} F(\boldsymbol{x}, u(\boldsymbol{x}), \mathrm{grad}[u(\boldsymbol{x})])\,\mathrm{d}\Omega - \int_{\Gamma^{\mathrm{N}}} \alpha(\boldsymbol{x})u(\boldsymbol{x})t^{\mathrm{p}}(\boldsymbol{x})\,\mathrm{d}\Gamma \tag{5.8}$$

and the scalar function $\alpha(\boldsymbol{x})$ is same as before, equation (5.3).

We now use this variational principle to derive a new weak formulation for the advective-diffusive system. We later show that for constant coefficients and homogeneous Dirichlet boundary conditions, this formulation is same as the optimal artificial diffusion method.

**Remark 5.1.** *It is well-known that, using the least-squares method one can construct a minimization problem for a given partial differential equation of the form $\mathcal{L}u - f = 0$ as follows:*

$$\underset{u(x)}{\mathrm{minimize}} \int_{\Omega} \|\mathcal{L}u - f\|^2 \mathrm{d}\Omega \tag{5.9}$$

But it should be noted that the Euler-Lagrange equation of the above minimization problem need not be the original differential equation $\mathcal{L}u - f = 0$. For example, if we consider the differential equation to be $d^2u/dx^2 = 0$ in $\Omega = (0,1)$, the corresponding minimization problem based on the least-squares method takes the following form:

$$\text{(5.10)} \qquad \text{minimize} \int_0^1 \left(\frac{d^2u}{dx^2}\right)^2 \, dx$$

However, the Euler-Lagrange equation of the above minimization problem is $d^4u/dx^4 = 0$ (which is not same as the differential equation we considered). Also, another difference between the minimization problem based on the least-squares method and the variational problem is in the regularity requirements. For example, in the minimization problem based on the least-squares method (5.10), $u$ should be twice differentiable (or, more precisely, $u \in H^2((0,1))$). On the other hand, the (standard) variational principle for the differential equation $d^2u/dx^2 = 0$ requires that $u$ be differentiable once (that is, $u \in H^1((0,1))$).

5.1. **A new stable formulation.** A necessary condition that the extremum of the functional $\mathcal{I}(u)$ has to satisfy is

$$\text{(5.11)} \qquad \left[\frac{d}{d\epsilon} \mathcal{I}(u(\boldsymbol{x}) + \epsilon w(\boldsymbol{x}))\right]_{\epsilon=0} = 0 \quad \forall w(\boldsymbol{x})$$

For convenience let us define the bilinear functional $\mathcal{B}(w; u)$ as

$$\mathcal{B}(w; u) := \left[\frac{d}{d\epsilon} \mathcal{I}(u(\boldsymbol{x}) + \epsilon w(\boldsymbol{x}))\right]_{\epsilon=0} = \int_\Omega \alpha(\boldsymbol{x}) \text{grad}\,[w(\boldsymbol{x})] \cdot (\boldsymbol{k}(\boldsymbol{x}) \text{grad}\,[u(\boldsymbol{x})]) \, d\Omega$$

$$\text{(5.12)} \qquad - \int_\Omega \alpha(\boldsymbol{x}) w(\boldsymbol{x}) f(\boldsymbol{x}) \, d\Omega - \int_{\Gamma^N} \alpha(\boldsymbol{x}) w(\boldsymbol{x}) t^{\text{p}}(\boldsymbol{x}) \, d\Gamma$$

A new weak formulation for the advective-diffusive system can be written as

$$\text{(5.13)} \qquad \text{Find } u(\boldsymbol{x}) \in \mathcal{U}_\alpha \text{ such that } \mathcal{B}(w; u) = 0 \; \forall w(\boldsymbol{x}) \in \mathcal{W}_\alpha$$

where $\mathcal{U}_\alpha$ and $\mathcal{W}_\alpha$ are weighted function spaces defined in equation (2.8) with weight function $\mu(\boldsymbol{x}) = \alpha(\boldsymbol{x})$. Recall that the scalar function $\alpha(\boldsymbol{x})$ is defined in equation (5.3). A corresponding finite element formulation can be written as

$$\text{(5.14)} \qquad \text{Find } u^h(\boldsymbol{x}) \in \mathcal{U}_\alpha^h \text{ such that } \mathcal{B}(w^h; u^h) = 0 \; \forall w^h(\boldsymbol{x}) \in \mathcal{W}_\alpha^h$$

where $\mathcal{U}_\alpha^h$ and $\mathcal{W}_\alpha^h$ are finite dimensional subspaces of $\mathcal{U}_\alpha$ and $\mathcal{W}_\alpha$, respectively. It is important to note the formulation (5.13) (and hence the formulation (5.14)) is valid even for spatial dimensions $d = 1, 2, 3$.

**Remark 5.2.** *Some notable differences between the bilinear functions $\mathcal{F}(w;u)$ and $\mathcal{B}(w;u)$ are as follows. A non-symmetric term similar to $\int_\Omega w(\boldsymbol{x})\boldsymbol{v}(\boldsymbol{x})\cdot\mathrm{grad}[u(\boldsymbol{x})]\,d\Omega$ is not present in the bilinear functional $\mathcal{B}(w;u)$. The bilinear functional $\mathcal{B}(w;u)$ is symmetric in the sense that*

$$\left[\frac{d}{d\epsilon}\,\mathcal{B}(w;u+\epsilon\bar{w})\right]_{\epsilon=0} = \left[\frac{d}{d\epsilon}\,\mathcal{B}(\bar{w};u+\epsilon w)\right]_{\epsilon=0}$$

*which is not the case for the bilinear functional $\mathcal{F}(w;u)$. The symmetry of $\mathcal{B}(w;u)$ (which, according to Vainberg's theorem, is necessary and sufficient for the existence of a scalar functional) should not be surprising as we defined the bilinear functional $\mathcal{B}(w;u)$ (5.12) as a directional derivative of a scalar functional. Another notable difference is that the bilinear functional $\mathcal{F}(w;u)$ is based on the standard $L^2$ inner product, which has the weight function (or measure density) to be 1. The bilinear functional $\mathcal{B}(w;u)$ is based on the weighted $L^2$ inner product with the weight function equal to $\alpha(\boldsymbol{x}) > 0$ (which is defined in equation (5.3)).*

**5.2. Application to the one-dimensional problem.** We now apply the new formulation to the one-dimensional problem defined in Section 3, and compare the difference equation produced by this new stable formulation with the one produced by the optimal artificial diffusion method given by equation (3.9) for an intermediate node. For the case at hand, the coefficients $v$ and $k$ are constants, and hence $\alpha(x) = \exp[-vx/k]$. The variable $u(x)$ is interpolated as

(5.15) $$u(x) = u_{j-1}N_{-1}(x) + u_j N_0(x) + u_{j+1}N_{+1}(x) \quad x_{j-1} \le x \le x_{j+1}$$

where the shape functions are defined as

(5.16a) $$N_{-1}(x) = \begin{cases} -\frac{(x-x_j)}{h} & -h \le x - x_j \le 0 \\ 0 & 0 < x - x_j \le +h \end{cases}$$

(5.16b) $$N_0(x) = \begin{cases} \frac{(x-x_j)+h}{h} & -h \le x - x_j \le 0 \\ \frac{h-(x-x_j)}{h} & 0 < x - x_j \le +h \end{cases}$$

(5.16c) $$N_{+1}(x) = \begin{cases} 0 & -h \le x - x_j \le 0 \\ \frac{(x-x_j)}{h} & 0 < x - x_j \le +h \end{cases}$$

The weighting function is also similarly interpolated. The difference difference equation produced by the new stable formulation at an intermediate node can be written as

(5.17) $$\gamma_{-1}u_{j-1} + \gamma_0 u_j + \gamma_{+1}u_{j+1} = f$$

The coefficients $\gamma_{-1}$, $\gamma_0$ and $\gamma_{+1}$ can be written as

$$\text{(5.18a)} \quad \gamma_{-1} = \frac{\int_{-h}^{+h} \alpha(x) k(x) N_0'(x) N_{-1}'(x) \, dx}{\int_{-h}^{+h} \alpha(x) N_0(x) \, dx}$$

$$\text{(5.18b)} \quad \gamma_0 = \frac{\int_{-h}^{+h} \alpha(x) k(x) \left(N_0'(x)\right)^2 \, dx}{\int_{-h}^{+h} \alpha(x) N_0(x) \, dx}$$

$$\text{(5.18c)} \quad \gamma_{+1} = \frac{\int_{-h}^{+h} \alpha(x) k(x) N_0'(x) N_{+1}'(x) \, dx}{\int_{-h}^{+h} \alpha(x) N_0(x) \, dx}$$

where a superscript prime denotes a derivative with respect to $x$. After simplification, the coefficients can be written as

$$\text{(5.19)} \quad \gamma_{-1} = -\frac{v}{2h}\left(1 + \coth(P_e^h)\right), \ \gamma_0 = \frac{v}{h}\coth(P_e^h), \ \gamma_{+1} = \frac{v}{2h}\left(1 - \coth(P_e^h)\right)$$

As one can see, the coefficients $\gamma_{-1}$, $\gamma_0$ and $\gamma_{+1}$ are, respectively, the same as the coefficients $\beta_{-1}$, $\beta_0$ and $\beta_{+1}$ (see equations (3.7) and (3.8)), which are obtained using the optimal artificial diffusion method. In Figure 2, we compare the numerical solutions obtained using the new stable formulation with the analytical solutions. As predicted by the theory, the new stable method produces nodally exact solution for all Péclet numbers. Both the theoretical and numerical studies have shown that the stable weak formulation proposed in the previous section is equivalent to the optimal artificial diffusion method. Hence, we have provided a variational basis for the artificial diffusion method, which has academic importance, and also has been the basis in developing and testing many stabilized formulations.

Although the new stable formulation is valid even in two- and three-dimensions, preliminary numerical studies have indicated that this formulation may not be practically feasible for large-scale problems due to the following reasons: evaluating the exponential function is expensive, one may require more Gauss points per element in order to evaluate integrals in the weak form, and also the resulting stiffness matrix will be ill-conditioned without an appropriate preconditioner. However, the new stable formulation, as shown in this paper, does have theoretical significance. It also highlights a possible route of developing robust stable formulations. This paper also highlights the relevance of the theoretical studies by Tonti on variational principles (for example, Reference [14]) to develop new stabilized formulations.

## 6. CONCLUSIONS

It is well-known that the classical Galerkin formulation (which is based on the standard weighted residual method) produces spurious node-to-node oscillations for the advection-diffusion equation for advection-dominated problems. For self-adjoint operators (e.g., the Laplacian equation), the standard weighted residual formulation can be obtained as a directional derivative of a scalar functional. However, for the advection-diffusion equation (using Vainberg's theorem) it can be shown that the standard weighted residual method cannot be obtained as a directional derivative of a scalar functional.

In this paper, we presented a variational principle for an advective-diffusive system, and derived a stable weak formulation. This resulting weak formulation when applied to one-dimensional problems gives rise to the same difference equation as the optimal artificial diffusion method, which produces exact nodal solutions when applied to one-dimensional problems with constant coefficients and forcing function. Hence, we have provided a variational basis for the optimal artificial diffusion method, which has been the cornerstone in developing many stabilized methods. We presented representative numerical results to corroborate our theoretical findings

## ACKNOWLEDGMENTS

The first author (K. B. Nakshatrala) acknowledges the financial support given by the Texas Engineering Experiment Station (TEES). The second author (A. J. Valocchi) was supported by the Department of Energy through a SciDAC-2 project (Grant No. DOE DE-FCO207ER64323). The opinions expressed in this paper are those of the authors and do not necessarily reflect that of the sponsors.

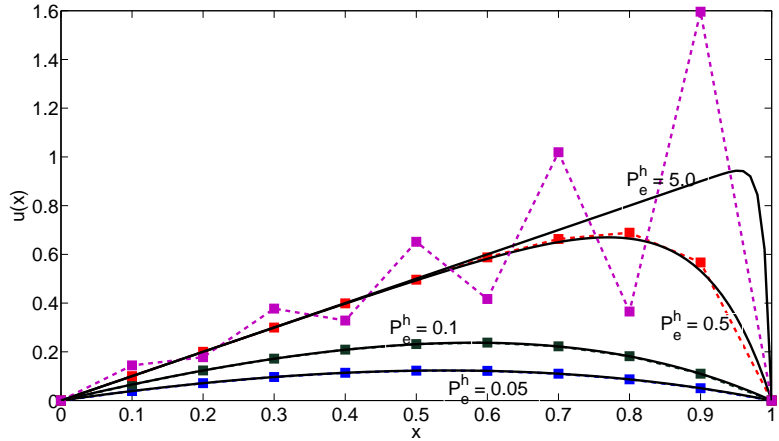

FIGURE 1. In this figure, we compare the Galerkin formulation (which is denoted using solid squares and dotted lines) with the analytical solution (which is denoted using solid continuous lines) for one-dimensional advection-diffusion equation with homogeneous Dirichlet boundary conditions for various $v/k$ ratios. We have taken the forcing function to be unity (i.e., $f = 1$), and $h = 0.1$. As expected, the Galerkin formulation produced spurious node-to-node oscillations for high Péclet numbers.

Correspondence to: Kalyana Babu Nakshatrala, Department of Mechanical Engineering, 216 Engineering/Physics Building, Texas A&M University, College Station, Texas-77843, USA. Tel: +1-979-845-1292

*E-mail address*: `knakshatrala@tamu.edu`

Albert Valocchi, Department of Civil and Environmental Engineering, University of Illinois at Urbana-Champaign, Urbana, Illinois-61801, USA.

*E-mail address*: `valocchi@uiuc.edu`

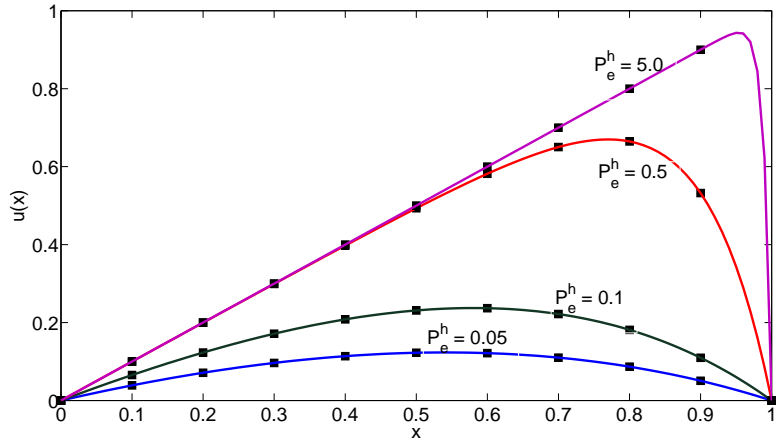

FIGURE 2. In this figure, we compare the numerical solution using the new stable formulation (which is denoted using solid squares) with the analytical solution (which is denoted using solid continuous lines) for one-dimensional advection-diffusion equation with homogeneous Dirichlet boundary conditions for various $v/k$ ratios. We have taken the forcing function to be unity (i.e., $f = 1$), and $h = 0.1$. As predicted by the theory, the new stable formulation produces nodally exact solutions for the chosen one-dimensional problem for all Péclet numbers. The continuous line denotes the analytical solution, and solid squares denote the numerical solution from the new stable formulation.